\documentclass[aps,prl,reprint,groupedaddress]{revtex4-1}

\usepackage{graphicx}
\usepackage{dcolumn}
\usepackage{bm}
\usepackage{amsmath}
\usepackage{hyperref}

\begin{document}

%\preprint{}

\title{Photon-Assisted Tunneling in a Biased Strongly Correlated Bose Gas}


\author{Ruichao Ma}
\author{M. Eric Tai}
\author{Philipp M. Preiss}
\author{Waseem S. Bakr}
\author{Jonathan Simon}
\author{Markus Greiner}

\email[]{greiner@physics.harvard.edu}
\affiliation{Department of Physics, Harvard University, Cambridge,
Massachusetts 02138, USA}

%\thanks{}


\date{\today}

\begin{abstract}
We study the impact of coherently generated lattice photons on an atomic Mott insulator subjected to a uniform force. Analogous to an array of tunnel-coupled and biased quantum dots, we observe sharp, interaction-shifted photon-assisted tunneling resonances corresponding to tunneling one and two lattice sites either with or against the force, and resolve multiorbital shifts of these resonances. By driving a Landau-Zener sweep across such a resonance, we realize a quantum phase transition between a paramagnet and an antiferromagnet, and observe quench dynamics when the system is tuned to the critical point. Direct extensions will produce gauge fields and site-resolved spin flips, for topological physics and quantum computing.
\end{abstract}

%\pacs{}
%\keywords{}

\maketitle

Advances in low-dissipation materials have enabled studies of coherent dynamics in strongly correlated many-body systems \cite{Stormer1999,Bloch2008}. Among other intriguing features, such systems may be excited to and studied in metastable states far from equilibrium. Such excitation takes a variety of forms, but is often realized by photon-assisted tunneling, where external photons provide the requisite excitation energy to drive spatial reorganization into long-lived excited states. This has been achieved via microwaves in coupled quantum dots \cite{Sakaki1994,Kouwenhoven1998} and Josephson junctions \cite{Shapiro1963}, terahertz radiation in semiconductor superlattices \cite{Rodwell1995}, and near-single-cycle pulses in condensed matter systems \cite{Wall2011}.

Cold atomic gases trapped in optical lattices provide an ideal playground for studying coherent, far-from-equilibrium dynamics. These systems mimic the solid state but are essentially defect- and dissipation-free. Accordingly, they offer a pristine platform for photon-assisted control \cite{Arimondo2008,Tino2008,Nagerl2010} and probing \cite{Jordens2008,Clement2009,Greif2011,Ernst2010} of ordering. The photons are usually introduced by classical modulation of either the lattice phase \cite{Arimondo2008,Tino2008,Tino2009} or amplitude \cite{Tino2011,Nagerl2010}. These engineered photons have been used to generate large-scale, coherent super-Bloch oscillations \cite{Nagerl2010,Arimondo2008,Tino2008} and further employed for precision measurement of applied forces \cite{Tino2011,Carusotto2005}. Weak interparticle interactions have been shown to act as a decoherence channel that damps  Bloch (and super-Bloch) oscillations and may be tuned away with a Feshbach resonance \cite{Nagerl2008,Fattori2008}. Modulation with additional spatial variation provides a promising route to gauge fields \cite{Kolovsky2011,Lukin2005} and other exotic topological effects \cite{Kitagawa2010}.

In line with the current effort to study strongly interacting materials via quantum simulation in cold atomic gases, there is a growing drive to investigate photon-assisted tunneling in this regime. Modulation spectroscopy presently provides the best temperature measurement of fermionic Mott insulators in the approach to quantum magnetism in the Fermi-Hubbard model\cite{Greif2011}. Strong photon dressing may be used to null and even negate tunneling \cite{Holthaus1992}, a technique that has been employed to drive the superfluid-Mott insulator transition \cite{Arimondo2009} and to simulate classical magnetism \cite{Sengstock2011}. Coherent control of cotunneling was recently demonstrated by driving an optical double well \cite{YuAo2011}. In this work we employ our high-resolution quantum gas microscope \cite{Bakr2009} to study photon-assisted tunneling in a scalable many-body system comprising a tilted, low-entropy Mott insulator \cite{Bakr2010,Sherson2010}. We explore both the occupation-number sensitivity of the tunneling resonance and the coherent many-body dynamics resulting from Landau-Zener sweeps of the photon energy.

\begin{figure}
    \includegraphics[width=0.41\textwidth]{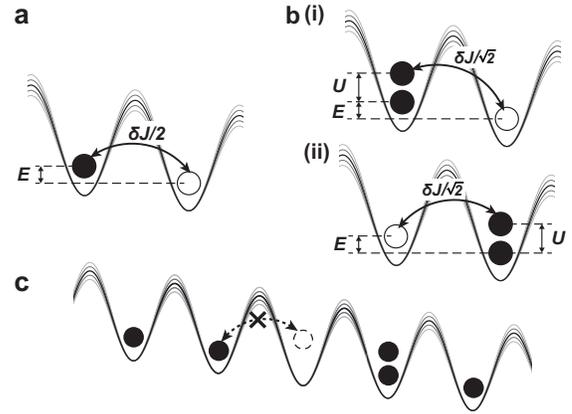}
    \caption{Photon-assisted tunneling. (a) For a single atom in a biased double well, tunneling is induced by lattice depth modulation at a frequency given by the energy gap $E$. (b) When there is one atom in each well, the energy gaps become $U+E$ for the right atom to tunnel onto the left atom ($\romannumeral 1$) and $U-E$ for tunneling in the other direction ($\romannumeral 2$). The effective tunneling under modulation is Bose-enhanced by a factor of $\sqrt2$. (c)The double well physics may be extended to a one-dimensional lattice at unity filling (e.g. $n=1$ Mott insulator). There is then an interaction blockade preventing photon-assisted tunneling on adjacent sites due to the resulting energy mismatch. Arrows denote photon-assisted tunneling, with open (closed) circles denoting the initial (final) location of the atom.}

     \label{setupfig}
\end{figure}

To understand photon-assisted tunneling in a lattice, it is instructive to begin with bosonic atoms in a biased, tunnel-coupled double well with time-dependent tunneling rate $J(t)$, bias $E$ between the left and right wells, and on-site interaction $U$ (Figs.~\ref{setupfig}a,b). The Hamiltonian for this system is given by:
\begin{equation*}
H=-J(t)(a_l^\dagger a_r+a_r^\dagger a_l)+\frac{E}{2}(a_l^\dagger a_l-a_r^\dagger a_r)+\frac{U}{2}(a_l^{\dagger2} a_l^2+a_r^{\dagger2} a_r^2)
\end{equation*}
Here $a_l^\dagger$ ($a_r^\dagger$) is the bosonic creation operator for the left (right) well. In the simple limit of  large bias $E\gg |J(t)|$, no modulation ($J(t)=J$), and one atom initially localized in the right well, tunneling populates the left well off-resonantly and hence with low probability $P_l(t)\approx\frac{4J^2}{E^2}$. By modulating the tunneling rate at the bias frequency $E$, $J(t)=J+\delta J \cos{E t}$, population is resonantly transferred between the two wells (Fig.~\ref{setupfig}a), resulting in full-amplitude Josephson-like tunneling oscillations occurring with an effective Rabi-frequency given by $\delta J/2$. We thus say that the modulation provides a photon to the atom with the requisite energy to tunnel.

If instead we begin with two strongly interacting ($U\gg|J(t)|$) atoms, one on each site of the double well, the energy cost to tunnel now becomes $U+E$ ($U-E$) for the atom in the right (left) well to tunnel to the left (right) well (Fig.~\ref{setupfig}b.~i,ii). By modulating at these new frequencies, one can induce tunneling in one direction or the other. Because of the indistinguishability of the atoms, the effective tunneling rate is Bose-enhanced to $\delta J/\sqrt{2}$.

The double well picture can be directly extended to atoms localized in a one-dimensional optical lattice (Fig.~\ref{setupfig}c).
As in the double well, tunneling is strongly suppressed in a Mott insulator by the repulsive onsite interactions \cite{Greiner2002}. A constant force resulting from an energy shift per lattice site of $E\neq U$ produces a tunneling energy gap that is dependent on the tunneling direction. Modulating the depth of the lattice (and hence, primarily, the tunneling rate) at a frequency equal to the energy gap enables the atoms to tunnel resonantly onto their neighbors. After such a photon-assisted tunneling event, two adjacent lattice sites each initially containing $n$ atoms become a site with $n+1$ atoms and a site with $n-1$ atoms.\\

\textit{Modulation Spectroscopy in a Tilted Optical Lattice}---
Our experiments begin with a two-dimensional Mott insulator of $^{87}$Rb atoms in a square optical lattice in the $x-y$ plane, with lattice constant $a=680$ nm, as described in previous work \cite{Bakr2010,Simon2011}. At an initial depth of $45E_r$ in both lattice directions, tunneling between sites is negligible on the time scale of our experiment. Here the recoil energy is defined as $E_r = h^2/8ma^2$, where $h$ is Planck's constant and $m$ the mass of $^{87}$Rb. 

 \begin{figure}
 \includegraphics[width=0.48\textwidth]{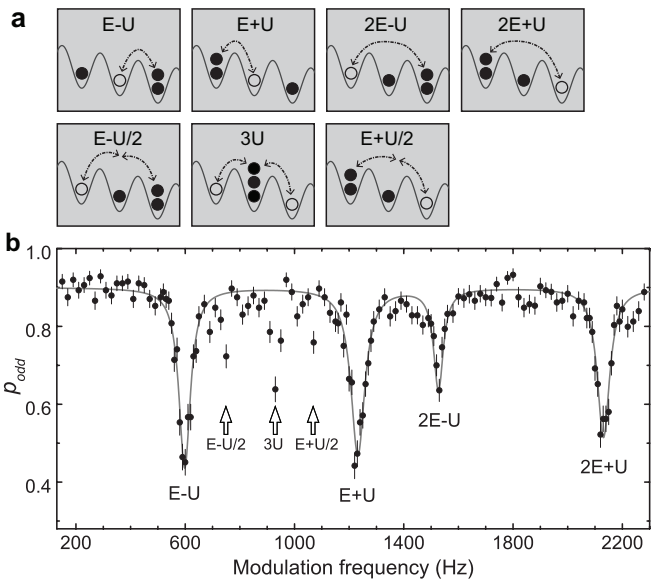}
 \caption{Modulation spectroscopy.  (a) Schematic of the observed tunneling processes.  (b) The occupation probability $p_{odd}$ versus the modulation frequency. $p_{odd}$ drops at each of the tunneling resonances, as many doublon-hole pairs are produced.  Each resonance is labeled in accordance with the corresponding process described in (a). The modulation of the $9E_r$ $x$-lattice by $\pm15\%$ corresponds to a Bose-enhanced photon-assisted tunneling rate of $2\pi\times 8$ Hz. $p_{odd}$ is averaged over a region of 25 lattice sites for 8 realizations. The solid curve is a four-Lorentzian fit to the $E\pm U$, $2E\pm U$ features. In this letter, all errors in $p_{odd}$ reflect 1$\sigma$ statistical uncertainties in the region averages.}
  \label{modspec}
 \end{figure}

A magnetic field gradient applied along the $x$ direction produces a lattice tilt of $48$ $\frac{\text{Hz/Site}}{\text{Gauss/cm}}$ for our Mott insulator, prepared in the ${|F=1,m_f=-1\rangle}$ state. To minimize tilt inhomogeneities we compensate the harmonic confining potential \cite{Simon2011}. We then adiabatically ramp down the lattice along $x$ to $9(1)E_r$, converting the system into uncoupled 1D chains with tunneling rate $J=2\pi\times30(7)$~Hz and measured onsite interaction $U=2\pi\times317(10)$ Hz, and apply a tilt of $E=2\pi\times915(10)$~Hz per lattice site. Modulation of the $x$-lattice depth produces photon-assisted tunneling, which is detected by \emph{in situ} fluorescence imaging with single-site resolution \cite{Bakr2009}. Our imaging technique is sensitive to the parity of site occupation \cite{Bakr2009}, resulting in a clear signature of the photon-assisted tunneling, which in a Mott insulator changes the occupation parity. The experiment is repeated under the same conditions in order to compute the probability $p_{odd}$ of odd occupation on each lattice site.

Figure~\ref{modspec} shows $p_{odd}$ versus modulation frequency in an $n=1$ Mott shell when the $x$-lattice is modulated by $\pm15\%$ for 500 ms, sufficient time for damping of the many-body oscillations. The principal features are peaks at $E\pm U$ which correspond to the creation of nearest neighbor doublon-hole pairs, and $2E\pm U$ which correspond to a second-order process creating next-nearest neighbor doublon-hole pairs. The dependence of the resonance locations on the tilt $E$ has been separately verified. We are also able to identify narrow resonances consistent with higher-order processes at $E\pm U/2$ and $3 U$, which require two assisted tunneling events. The principal peaks have typical widths of $\sim$60 Hz, set by residual lattice disorder specific to our apparatus \cite{Bakr2010}.\\
%\textit{In situ} imaging allows us to obtain high spectral resolution by focusing on homogeneous regions of the cloud. 

 \begin{figure}
 \includegraphics[width=0.48\textwidth]{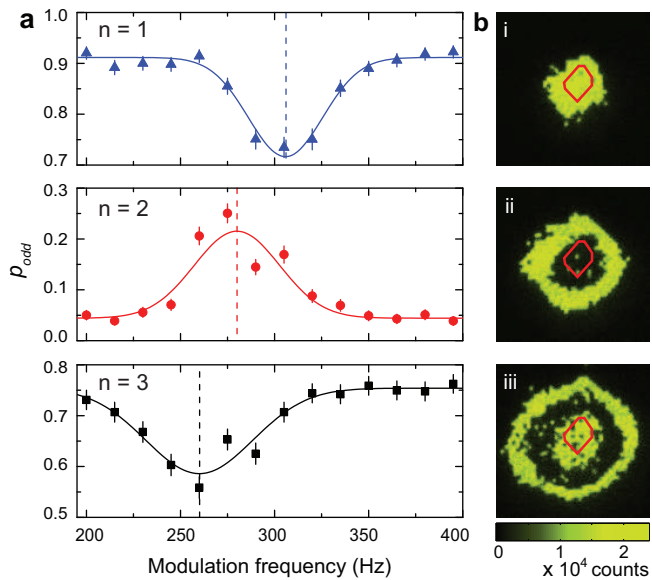}
 \caption{Number-sensitive photon-assisted tunneling using multiorbital shifts. (a) Modulation spectra of the $U-E$ resonant peaks in $n=1, 2, 3$ Mott insulator shells. Fitted to Gaussian profiles (solid curves), the peaks are located at $306(2)$~Hz, $280(3)$~Hz, and $260(3)$~Hz respectively (dashed lines). (b) \textit{In situ} images of the $n=1, 2, 3$ Mott shells ($\romannumeral 1 - \romannumeral 3$) without modulation, with the 49-site region studied in (a) enclosed by the red solid line.}
 \label{multiorb}
 \end{figure}
 
\textit{Occupation-Sensitive Photon-Assisted Tunneling}---Taking advantage of the high spectral and spatial\cite{Bakr2009} resolution achievable in an optical lattice, we demonstrate occupation-sensitive photon-assisted tunneling by making use of the multiorbital shift \cite{Will2010,Porto2009}. This shift arises due to virtual transitions to higher bands, producing a total onsite interaction energy $U_{\text{MO}}(n)$ for a site with $n$ atoms, from effective multibody interactions $U_n$:
\begin{align*}
	U_{\text{MO}}(n) & = \frac{U_2}{2!}n(n-1) + \frac{U_3}{3!}n(n-1)(n-2) \\
						  &+ \frac{U_4}{4!}n(n-1)(n-2)(n-3)+\ldots
\end{align*}

For the Mott insulator shell with $n$ atoms per site, the $U-E$ resonance corresponds to the conversion of $n$ atoms on each of two adjacent sites to $(n-1)$ and $(n+1)$. The energy cost of this process is thus $\delta U_{\text{MO}}=U_{\text{MO}}(n+1)+U_{\text{MO}}(n-1)-2U_{MO}(n)-E$. The photon-assisted tunneling resonances of the $n=2$ and $n=3$ shells are therefore shifted by $\delta^{n=2}_{\text{MO}}=U_3$ and $\delta^{n=3}_{\text{MO}}=2 U_3+U_4$ relative to that of the $n=1$ resonance. A multiband calculation \cite{Porto2009} predicts $U_3=2\pi\times -23(1)$ Hz and a variational calculation (over Gaussian wavepacket r.m.s. size) yields $U_4=2\pi\times 6(1)$ Hz for our experimental parameters, with the errorbars arising from uncertainty in the measured $U_2=429(15)$ Hz (in an $18 E_r$ $x$-lattice).

Figure~\ref{multiorb} shows a shell-resolved measurement of the tunneling resonance frequency for tilt $E=2\pi\times120(15)$ Hz. We observe shifts of $-26(5)$ Hz and $-46(5)$ Hz for the $n=2$ and $n=3$ shells relative to the $n=1$ shell.  The theory, discussed above, predicts $-23(1)$ Hz and $-40.3(1)$ Hz, respectively.  The imperfect agreement for the $n=3$ shell likely arises from higher-order effects such as superexchange interactions between adjacent lattice sites, which are not included in the aforementioned model and become increasingly important for larger occupations.\\

\textit{Quantum Magnetism}---While the modulated tunneling is initially resonant at \textit{every} site, once an atom has tunneled both of its neighbors are blocked by the resulting energy gap (Fig.~\ref{setupfig}). As recently demonstrated experimentally\cite{Simon2011}, this many-body system maps directly onto a one-dimensional quantum Ising model \cite{Sachdev2002}. Modulation enables us to rapidly control the longitudinal field $h_z$ by varying the modulation frequency $\omega_{mod}$.

\begin{figure}
 \includegraphics[width=0.48\textwidth]{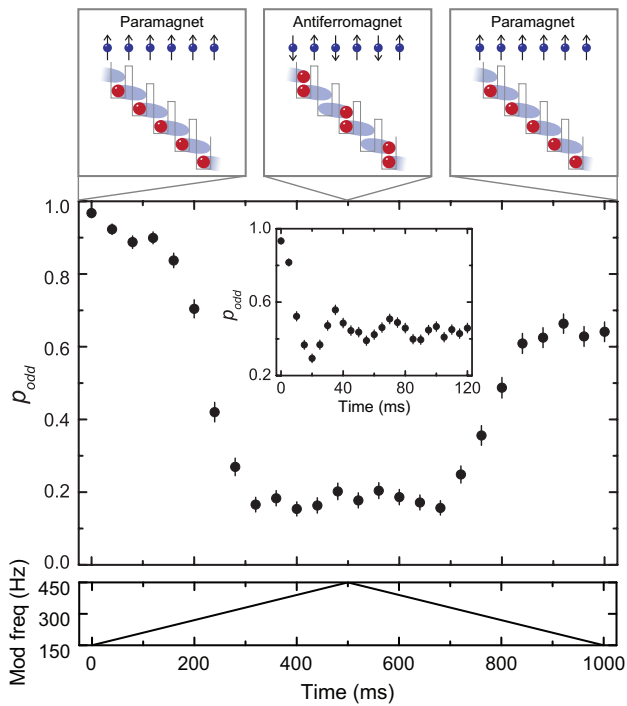}
 \caption{Quantum magnetism by lattice modulation. For an $n=1$ Mott shell tilted by $E=2\pi\times710(25)$ Hz, the modulation frequency is chirped linearly from 150 Hz to 450 Hz and back in a total of 1 second, thus tuning through the $|U-E|$ resonance to produce doublon-hole pairs and then back to restore the double-hole pairs to singly-occupied sites (illustrated above). The magnetic model \cite{Simon2011} maps this to a quantum phase transition from the paramagnet to an antiferromagnet and back. The lattice depths are $18E_r$ and $45 E_r$ in the $x$- and $y$- directions, respectively, corresponding to an onsite interaction energy of $U=2\pi\times 416(15)$ Hz. The $x$-lattice depth is modulated by $\pm70\%$, producing a Bose-enhanced photon assisted tunneling rate of $2\pi\times 7.4$ Hz. Inset: Dynamics of a many-body quench, produced by driving at $\omega_{mod}=E-U$, for $E=2\pi\times890(10)$ Hz and $U=2\pi\times317(10)$ Hz, for a reduced $x$-lattice depth of $9 E_r$. The photon-assisted tunneling rate (with $\pm17\%$ lattice depth modulation) is $2\pi\times 8.1$~Hz.}
\label{MagRamp}
\end{figure}

We drive a quantum phase transition in this magnetic model by performing a Landau-Zener sweep of the modulation frequency across the $|U-E|$ resonance of the $n=1$ shell, and subsequently \textit{back}, adiabatically creating and destroying doublon-hole pairs. Figure~\ref{MagRamp} shows such a sweep, demonstrating a quantum phase transition between paramagnetic ($p_{odd}=1$) and antiferromagnetic ($p_{odd}=0$) many-body ground states.  The $\sim80$\% conversion into doublon-hole pairs is limited by atom loss (due to the long $1$ second sweep), noise on the lattice tilt, and residual lattice disorder.

The rapid tunability of the lattice modulation enables us to perform a nearly instantaneous quench of the magnetic model to the vicinity of the critical point $\omega_{mod}=E-U$. The subsequent dynamics are shown in the inset in Fig. ~\ref{MagRamp}, and exhibit oscillations which are damped by the many-body blockade\cite{Sengupta2004}, as well as residual lattice disorder. This approach provides substantial added flexibility, as the phase, amplitude, and frequency of the tunneling may be arbitrarily controlled via the modulation parameters.

In summary, we have performed an \textit{in situ} study of photon-assisted tunneling in a Mott insulator by amplitude modulation of a tilted optical lattice. We have realized occupation-sensitive control of tunneling by making use of multiorbital effects, with immediate applications in the generation of low-dimensional anyons \cite{Keilmann2010}.
The reversibility of the modulation-driven tunneling is demonstrated in a many-body setting that can be used to study magnetic models in optical lattices \cite{Simon2011,Sachdev2002}, with promising extensions to higher dimensions and other lattice geometries \cite{Pielawa2011}. The high resolution afforded by the quantum gas microscope should enable locally controlled photon-assisted tunneling, with applications in quantum computing \cite{Schneider2011} and quantum Hall physics \cite{Kolovsky2011,Lukin2005}. Tunable second-neighbor couplings could be engineered through multichromatic modulation schemes, while the dynamics of such systems could be driven and studied by time-dependent tuning of the modulation parameters.

\begin{acknowledgments}
We would like to thank Andrew Daley, Eugene Demler and Stefan Trotzky for interesting discussions. This work was supported by grants from the Army Research Office with funding from the DARPA OLE program, an AFOSR MURI program, and by grants from the NSF.
\end{acknowledgments}

\bibliographystyle{apsrev4-1}
\end{document}